\documentclass[runningheads]{svmult}

\usepackage{makeidx}   
\usepackage{graphicx}  
\usepackage{subeqnar}  
\usepackage{multicol}  
\usepackage{cropmark}  
\usepackage{physprbb}  

\newcommand{\greeksym}[1]{{\usefont{U}{psy}{m}{n}#1}}

\newcommand{\uGamma}{\mbox{\greeksym{G}}}
\usepackage{lscape}

\begin{document}
\title*{Active Galactic Nuclei at the Crossroads of Astrophysics}
%
%
%
%
\titlerunning{AGN at the Crossroads of Astrophysics}
%
\author{Andrei Lobanov
\and J. Anton Zensus}

\authorrunning{Lobanov \& Zensus}
%
%
\institute{Max-Planck-Institut f\"ur Radioastronomie, Auf dem H\"ugel 69, 53121 Bonn, Germany}

\maketitle              

\begin{abstract}
Over the last five decades, AGN studies have produced a number of
spectacular examples of synergies and multifaceted approaches in
astrophysics. The field of AGN research now spans the entire
spectral range and covers more than twelve orders of magnitude in the
spatial and temporal domains. The next generation of astrophysical
facilities will open up new possibilities for AGN studies,
especially in the areas of high-resolution and high-fidelity imaging
and spectroscopy of nuclear regions in the X-ray, optical, and radio
bands. These studies will address in detail a number of critical
issues in AGN research such as processes in the immediate vicinity
of supermassive black holes, physical conditions of broad-line and
narrow-line regions, formation and evolution of accretion disks and
relativistic outflows, and the connection between nuclear activity and
galaxy evolution. 
\end{abstract}

\section{Nuclear activity in galaxies}

Recent years have witnessed substantial progress in studies of
active galactic nuclei (AGN).  Activity is observed in galaxies
throughout the entire range of the electromagnetic
spectrum~\cite{lobanov:fab99}. It is manifested on spatial and temporal
scales that span over 12 orders of magnitude, ranging from several
Schwarzschild radii to megaparsecs and from minutes to millions of
years~\cite{lobanov:pet02}. On the surface, the activity of galactic
nuclei comes in many faces, outlined by the uncomfortably large number
of the AGN classes --- yet the underlying mechanism and physical
conditions in the nuclear regions of all galaxies are probably more
similar than  may appear at first
glance~\cite{lobanov:ant93,lobanov:gm98,lobanov:ver00}.  Last but not
least, AGN are now detected at redshifts of 6 and
beyond~\cite{lobanov:fan+03}, and they may be closely connected to the
formation and evolution of the large-scale
structure~\cite{lobanov:mb00} and to the epoch of reionization of the
Universe~\cite{lobanov:fan+02,lobanov:mad+04,lobanov:ric+04}.

There is growing acceptance of ubiquity of the nuclear activity in
galaxies and its connection to the presence of supermassive black
holes (SMBH) in the centres of all massive
galaxies~\cite{lobanov:and+04,lobanov:gh04,lobanov:ho99a,lobanov:ho99b}. Observations
in the infrared~\cite{lobanov:gc00} and sub-millimetre
bands~\cite{lobanov:mb00} indicate that almost every galaxy exhibits
certain characteristics previously thought to be related only to a
much narrower class of powerful AGN. This implies that virtually all
massive galaxies may go through a strong AGN phase in the course of
their cosmological evolution.  Historically, observational studies of
AGN have been concentrated largely within six broadly defined areas.

\emph{1.~Surveys} have been used effectively, first in the optical and
radio domains and later throughout much of the electromagnetic
spectrum, to find correlations between different types of AGN and
to trace their cosmological evolution. The databases provided by the
ROSAT, IRAS, ISO, SIRTF and VLA (FIRST~\cite{lobanov:bwh95} and
NVSS~\cite{lobanov:con+98}) surveys have become the true cornerstones
of AGN research, as well as the HST and Chandra deep field
observations. These efforts will be taken further by the ongoing SDSS
survey~\cite{lobanov:aba+03} and the Spitzer
surveys~\cite{lobanov:fad+04}.  The Spitzer data have already yielded a
fundamental result on the obscuration in AGN, indicating that as much
as three quarters of all AGN are likely to be obscured at all
redshifts~\cite{lobanov:urr04}, which implies that the number of active
galaxies may be much higher than was thought. AGN studies in the
next decade will also benefit from multi-wavelength surveys such as
the great observatories origins deep survey
(GOODS~\cite{lobanov:gia+04}) which addresses AGN structure and
demographics~\cite{lobanov:hec+04}, unification scheme and and
AGN--SMBH co-evolution~\cite{lobanov:gm98,lobanov:urr04}. A number of
ground and space VLBI surveys undertaken in the radio in the
centimetre~\cite{lobanov:fc00,lobanov:fom+00,lobanov:hen+95,lobanov:hir+00,lobanov:kel+98,lobanov:lis03b,lobanov:pr88,lobanov:pol+95,lobanov:tay+94,lobanov:tay+96,lobanov:tha+95,lobanov:xu+95}
and millimetre~\cite{lobanov:lob+00,lobanov:ldp98,lobanov:rbb98} domains
have revealed a wealth of information about
morphology~\cite{lobanov:hor+04,lobanov:kov+05,lobanov:zen+02},
kinematics~\cite{lobanov:kel+04} and evolution of radio-emitting
material in the nuclear regions and relativistic outflows in AGN on
scales down to fractions of a parsec. VLBI surveys have also been used
effectively for addressing a variety of general astrophysical problems
including AGN evolution~\cite{lobanov:sha+96} and population
modelling~\cite{lobanov:lis03,lobanov:ls02}, jet
formation~\cite{lobanov:lob+00}, fundamental astrophysical emission
processes~\cite{lobanov:kel02,lobanov:lm99}, and
cosmology~\cite{lobanov:gkf99}.

\begin{figure}[t]
\begin{center}
\includegraphics[width=0.9\textwidth,angle=0,bb=14 14 326 253,clip=true]{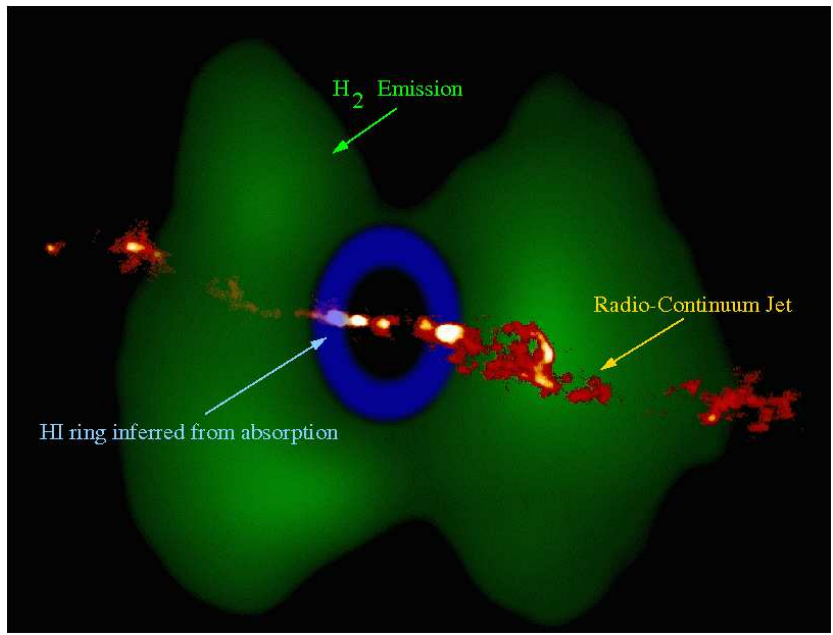}
\caption{Montage of the inner 250 pc of NGC 4151. The H$_2$ emission
traces a torus, the H{\,\scriptsize I} absorption comes from a ring
inside the torus. Ionized gas (black) is assumed to fill the torus
inside the H{\,\scriptsize I} ring~\cite{lobanov:mun+03}.}
\label{lobanov:fig1}
\end{center}
\end{figure}

\emph{2.~Morphological studies} of AGN have been done extensively in
the radio~\cite{lobanov:kas+93,lobanov:ohc89},
optical~\cite{lobanov:per+99} and X-ray
regimes~\cite{lobanov:mar+01,lobanov:mar+02}, recovering large-scale
structures produced by AGN.  Circumnuclear disks and tori have been
revealed conclusively in recent optical, near-infrared and radio
observations~\cite{lobanov:jaf+04,lobanov:kbg03,lobanov:mun+03,lobanov:wei+04}
(Fig.~\ref{lobanov:fig1}). Most spectacularly, recent X-ray images
obtained with Chandra have shown the shocks and ripples in the
intergalactic medium excited by the central engine in
NGC\,1275~\cite{lobanov:fab+03}.

\emph{3.~Variability studies} of long- and short-term changes of
continuum and line emission have enabled detailed investigations of
non-stationary processes in the immediate vicinity of the SMBH and
understanding better the physics of nuclear regions of
AGN~\cite{lobanov:ulr92}. Timescales probed by these studies range
from hours to several
decades~\cite{lobanov:aah03,lobanov:bvf03,lobanov:pet02,lobanov:sha+01,lobanov:sta+04,lobanov:wam+97,lobanov:zhe96}.
The variable continuum flux is believed to be responsible for ionizing
the cloud material in the broad-line region (BLR).  Optical spectral
line and continuum variability data have been combined to model the
BLR~\cite{lobanov:kas+00,lobanov:pet02,lobanov:pet+02}, which provided a
reliable method for estimating the mass of the central black holes in
a number of AGN~\cite{lobanov:pet02}. In radio-quiet AGN, short-term
variations of the continuum X-ray
emission~\cite{lobanov:are+05,lobanov:bri+04} are most likely connected
with processes occurring the accretion disk at distances of $\sim
10$--$100\,R_\mathrm{g}$ ($R_\mathrm{g} = G\,M_\mathrm{bh}/c^2$ is the
gravitational radius for a black hole of mass $M_\mathrm{bh}$, where
$G$ is the Newtonian gravitational constant) from the black
hole~\cite{lobanov:mdp93}. This conclusion is contested by
multi-wavelength monitoring
observations~\cite{lobanov:die+98,lobanov:pet02} showing no apparent
time delay between flux variations in the ultraviolet (UV) and optical
continua, which indicates that instabilities in the accretion disk or
random fluctuation in the accretion rate alone cannot explain the
continuum variability~\cite{lobanov:pet02}. In radio-loud AGN,
continuum emission from the relativistic plasma in the jet dominates
at all energies~\cite{lobanov:umu97,lobanov:wor05}, swamping the X-ray
emission associated with the accretion flow.  Hence, the continuum
variability in radio-loud AGN may be related to both the jet and the
instabilities of accretion flows~\cite{lobanov:mdp93,lobanov:umu97} near
the central engine.

Long-term variability of
X-ray~\cite{lobanov:mar+02,lobanov:min+03,lobanov:shi02} and radio
emission~\cite{lobanov:aah03} probably reflects instabilities
developing in the disk~\cite{lobanov:lr05} and shocks propagating in
the relativistic jet~\cite{lobanov:fer98}.  Ejections of new jet
components have been reported to occur after minima in the X-ray light
curve~\cite{lobanov:mar+02}, suggesting that the jet is fed by the
material falling onto the black hole from inner radii of the accretion
disk.  No observational evidence has been reported so far for a link
between optical/UV continuum variability and the compact radio jet.

Intraday variations observed in the
radio~\cite{lobanov:kra+03,lobanov:ww95}, optical~\cite{lobanov:hw96},
IR~\cite{lobanov:gup+04}, and X-ray domains~\cite{lobanov:are+05} most
likely reflect processes in the immediate vicinity of the central
engine and in internal shocks developing in relativistic
outflows~\cite{lobanov:ssp98,lobanov:ssp99,lobanov:spa+01}.  In this
respect, intraday variability in AGN may be similar to quasi-periodic
oscillations observed in Galactic X-ray binaries. Interstellar
scintillations are also likely to play a role in the intraday
variations observed in the radio
regime~\cite{lobanov:big+02,lobanov:rkd02}.

\emph{4.~Multifrequency campaigns}, albeit proven difficult to organize,
have yielded accurate broad-band spectra of
AGN~\cite{lobanov:mar+94,lobanov:tur+99} and have enabled detailed
investigations of physical processes governing the production of the
non-thermal continuum emission from circumnuclear regions in
AGN~\cite{lobanov:har+01,lobanov:tur+00}.  Multifrequency data were used
to determine the spectral energy distribution in AGN in the quiescent
and flaring stages~\cite{lobanov:fos+98,lobanov:mat04,lobanov:smu96} and
to characterize the properties of the synchrotron emission in
relativistic outflows~\cite{lobanov:deh05,lobanov:lob98b}.

\emph{5.~Spectroscopic studies} in the
optical~\cite{lobanov:bec+01,lobanov:gre+96,lobanov:mar+03,lobanov:whi+00}
infrared~\cite{lobanov:lut+00} radio~\cite{lobanov:lob05b}, and
increasingly also in the X-ray
domain~\cite{lobanov:fab+95,lobanov:fab+02,lobanov:mal+97,lobanov:nan+97a,lobanov:nan+97b,lobanov:ogl+01,lobanov:sam+98,lobanov:tan+95,lobanov:yaq+00}
have provided measurements of the gas and stellar kinematics near the
central black hole~\cite{lobanov:zak+03}, in the broad and narrow line
regions of galactic nuclei. These studies will be greatly enhanced
with the systematic spectroscopy provided by the SDSS data and with
the high-energy spectroscopy data from the existing and planned X-ray
missions.

\emph{6.~Dedicated monitoring programs} have been employed successfully
virtually in all observational domains to study in detail the dynamics
and physical conditions in the nuclear regions of a number of
prominent AGN.  Structure of the nuclear regions has been imaged
extensively with high resolution radio interferometry, uncovering the
evolution of radio emitting plasma on linear scales from $\approx
100R_\mathrm{g}$~\cite{lobanov:jbl99,lobanov:kri+02} to $\approx
1$\,kpc~\cite{lobanov:wii01,lobanov:zen97,lobanov:zen+96}. Radio
monitoring programs are now reaching timescales of several
decades~\cite{lobanov:cou98,lobanov:lz99,lobanov:lr05}. Radio frequency
observations of maser emission have been used to probe the presence of
accretion disks~\cite{lobanov:her+99,lobanov:miy+95} and molecular
tori~\cite{lobanov:kb04} around putative black holes in the centres of
AGN.  Extragalactic maser emission has provided direct evidence of
interaction between the dense molecular material and the ionization
cones~\cite{lobanov:gal+96} or nuclear
jets~\cite{lobanov:cla+98,lobanov:pec+03}.  Properties of relativistic
outflows have been connected to the physical conditions in the nuclear
regions~\cite{lobanov:kad+04,lobanov:lob98} and used for identifying
possible binary systems of SMBH in
AGN~\cite{lobanov:ca04,lobanov:lr05,lobanov:rom+00}.

\begin{figure}[t]
\begin{center}
\includegraphics[width=0.45\textwidth,bb=0 0 504 503,clip=true]{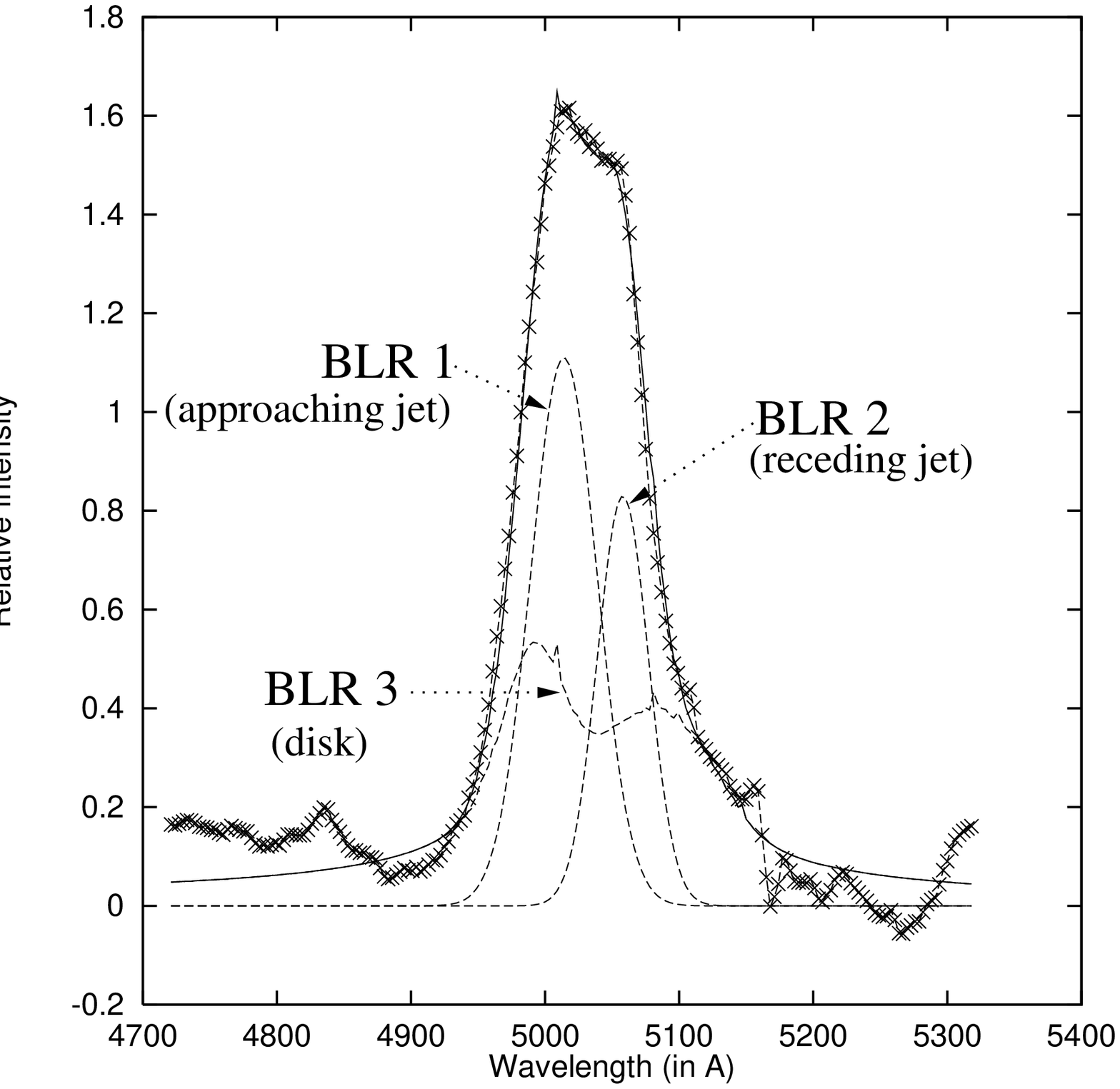}
\includegraphics[width=0.45\textwidth,bb=-94 37 691 805,clip=true]{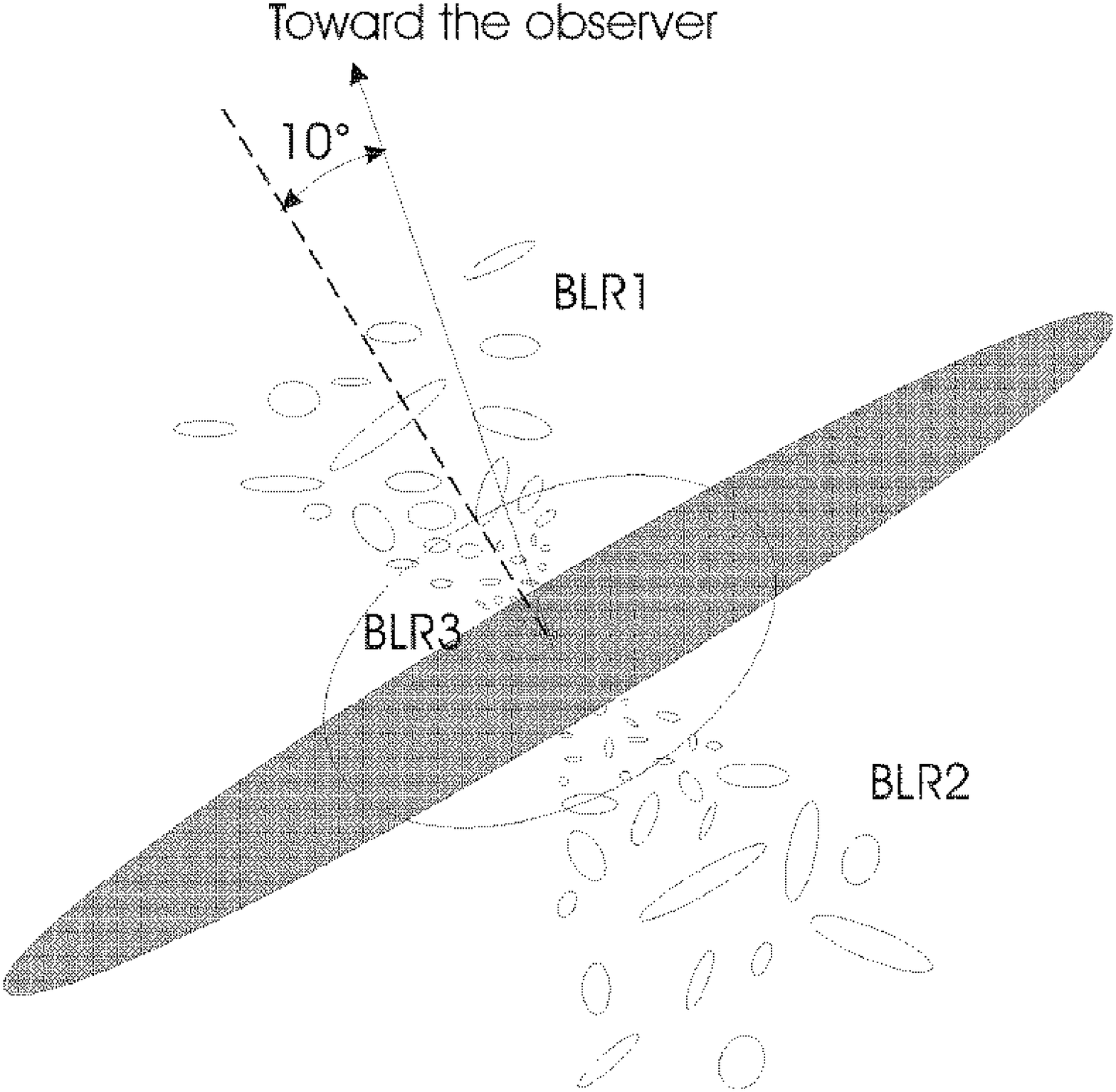}
\caption{{\bf Left}: The observed H$_\beta$ line of Ark\,120 fitted by
the two-component model (solid line): a disk (in the line wings) plus
an outflow-related component of the BLR (the core of the line). The
two components in the core correspond to the line emission generated
in the approaching (left) and receding (right) part of a bipolar
outflow.  {\bf Right}: The scheme of the proposed BLR model for
Ark\,120~\cite{lobanov:pop+01}.} 
\label{lobanov:fig2}
\end{center}
\end{figure}

High-resolution radio observations of Seyfert galaxies have offered an
opportunity to study the optical emission from the immediate vicinity
of the central engine of AGN, on scales comparable to the largest
extent of the BLR.  Optical observations of broad emission lines have
been instrumental for understanding the structure and dynamics of the
circumnuclear gas, and they have yielded an attractive concept of a
disk-like morphology of the BLR~\cite{lobanov:pop+04}. The detection
and modelling of several double-peaked Balmer lines has further
supported this idea. In most cases, the double-peaked emission lines
can be explained with line emission that originates in the
disk~\cite{lobanov:eh03}.  There are, however, two very notable
exceptions among the radio-quiet AGN. The core of the broad
double-peaked emission lines in RX\,J1042+1212~\cite{lobanov:pmc96} and
Ark\,120~\cite{lobanov:pop+01} is most likely to originate from an
outflow-related component of the BLR (Fig.~\ref{lobanov:fig2}). It is
possible that this component is indeed related to an interaction
between the relativistically moving plasma and high-velocity clouds
situated in a sub-relativistic outflow/wind surrounding the
jet~\cite{lobanov:elv00,lobanov:mc97,lobanov:psk00}. In this case, the
broad-line emitting region associated with the jet/outflow can be
located at a significant distance from the nucleus, perhaps as large
as $\sim 1$\,pc.  Verifying the existence of the outflow-related
component of the BLR and testing the putative connection between the
BLR and relativistic outflows in Seyfert galaxies can therefore be
viewed as critical experiments that may advance our understanding of
the nuclear activity in galaxies.

\section{A synthetic view of AGN}

The nuclear environment in active galaxies hosts a variety of physical
phenomena on scales ranging from $\sim$1--2\,$R_\mathrm{g}$ to $\sim
100$ parsecs. Table~\ref{lobanov:tb1} lists most important components
of the nuclear regions in galaxies. They cover over seven and ten
orders of magnitude in linear and temporal scales, respectively.  This
presents a serious challenge to any attempt of creating a single
framework for describing all aspects of nuclear activity. The key
constituents of the nuclear environment can be divided into six broad
categories:

\emph{Accretion disks and infalling material} producing strong
  continuum and line emission in the high-energy (X-ray and possibly
  even $\uGamma$-ray) regime~\cite{lobanov:are+05,lobanov:iwa+96}, and
  often exhibiting maser lines in the radio
  regime~\cite{lobanov:her+99}.
  
\emph{Broad-line and narrow-line regions} detected in the line
  emission and absorption in the optical regime~\cite{lobanov:pet02}.
  
  \emph{Obscuring torus} composed of clouds of gas and
  dust~\cite{lobanov:ant93} and manifested by emission and absorption,
  primarily in the optical~\cite{lobanov:vbd03}, infrared~\cite{lobanov:vbd04}
  and radio domains~\cite{lobanov:mun+03}.

\emph{Bipolar outflows} detected in continuum emission throughout much
  of the electromagnetic spectrum. Sub-relativistic outflows are also
  manifested by the optical absorption lines (BAL
  outflows~\cite{lobanov:elv00,lobanov:gal+04})
  
\emph{Nuclear stellar population} detected in the optical through
  near IR regimes via velocity dispersion or individual stellar proper
  motions in the nearest AGN in
  Sgr\,A$^\star$~\cite{lobanov:eck+02,lobanov:ghe+00}.

\emph{Secondary black holes} in multiple black hole systems, which can
be inferred from characteristic emission and structural variability in
the X-ray through radio
regimes~\cite{lobanov:ca04,lobanov:lr05,lobanov:roo85b}.

\begin{table}[t]
\caption{Characteristic scales in the nuclear regions in active galaxies}
\label{lobanov:tb1}
\begin{center}
\begin{tabular}{rccccc}\hline\hline
   & $l$ & $l_8$ & $\theta_\mathrm{Gpc}$ & $\tau_c$ & $\tau_\mathrm{orb}$ \\ 
  & [$R_\mathrm{g}$] & [pc] & [mas]& [yr] & [yr] \\ \hline
Event horizon:           &1--2          &$10^{-5}$           &$5\times 10^{-6}$    &0.0001     & 0.001 \\
Ergosphere:              &1--2          &$10^{-5}$           &$5\times 10^{-6}$    &0.0001     & 0.001 \\
Accretion disk:          &10$^1$--10$^3$&$10^{-4}$--$10^{-2}$&$0.005$              &0.001--0.1 & 0.2--15 \\
Corona:                  &10$^2$--10$^3$&$10^{-3}$--$10^{-2}$&$5\times 10^{-3}$    &0.01--0.1& 0.5--15 \\
Broad line region:       &10$^2$--10$^5$&$10^{-3}$--1        &$0.05$               &0.01--10   & 0.5--15000 \\
Molecular torus:         &$>$10$^5$     &$>$1                &$>$$0.5$             &$>$10      & $>$15000 \\
Narrow line region:      &$>$10$^6$     &$>$10               &$>$5                 &$>$100     & $>$500000 \\
Jet formation:           &$>$10$^2$     &$>$$10^{-3}$        &$>$$5\times 10^{-4}$ &$>$0.01    & $>$0.5 \\
Jet visible in the radio:&$>$10$^3$     &$>$$10^{-2}$        &$>$$0.005$           &$>$0.1     & $>$15 \\ \hline
\end{tabular}
\end{center}
{\bf Column designation:} $l$ -- dimensionless scale in units of the
gravitational radius, $G\,M/c^2$; $l_8$ -- corresponding linear scale,
for a black hole with a mass of $5\times 10^8\,$M$_{\odot}$;
$\theta_\mathrm{Gpc}$ -- corresponding largest angular scale at 1\,Gpc
distance; $\tau_c$ -- rest frame light crossing time;
$\tau_\mathrm{orb}$ -- rest frame orbital period, for a circular
Keplerian orbit.
\end{table}

\begin{figure}[t]
\begin{center}
\includegraphics[width=0.9\textwidth,angle=0,bb=14 14 326 249,clip=true]{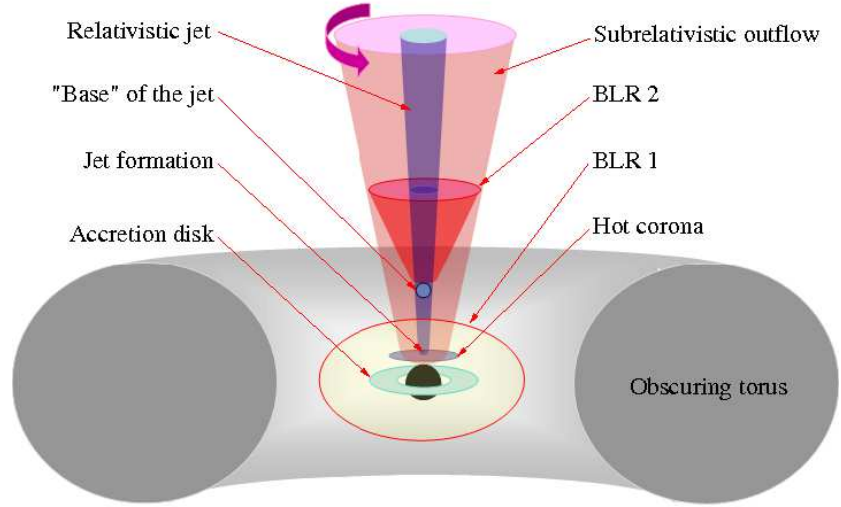}
\caption[]{A sketch of the nuclear region in an active galaxy (the
drawing is made not to scale and shows only the approaching jet).  The
broad-line emission is likely to be generated both near the disk
~\cite{lobanov:pet02} (an ellipsoidal BLR\,1, ionized by the emission from a
hot corona~\cite{lobanov:fab04,lobanov:pon+04} or the accretion disk~\cite{lobanov:fr93}) and
in a rotating subrelativistic outflow~\cite{lobanov:elv00,lobanov:mc97,lobanov:psk00}
surrounding the jet (a conically-shaped BLR\,2, ionized by the
emission from the relativistic plasma in the jet). BLR\,2 is evident
in the broad-line emission when the jet emission dominates the optical
continuum. BLR\,1 may be manifested in the broad-line emission when
the jet contribution to the ionizing continuum is small.}
\label{lobanov:fig3}
\end{center}
\end{figure}

The AGN studies outlined in the previous section have provided 
substantial knowledge about each of these aspects of nuclear activity.
This has enabled the construction of a synthetic, ``unified'' picture of
active galaxies, in which the entire spectrum of galactic activity is
described as an intricate interplay between the physical conditions
and orientation of individual constituents of an active
nucleus~\cite{lobanov:ant93,lobanov:gm98}. In the current AGN paradigm,
several key components, including the accretion disk, the broad-line
region and the relativistic outflow (jet), play major
roles~\cite{lobanov:fab99}, together with the putative central SMBH
(Fig.~\ref{lobanov:fig3}). A fraction of the infalling material forms
an outflow along the rotational axis of the
SMBH~\cite{lobanov:bp82,lobanov:bz76,lobanov:fer98,lobanov:mc97} and a
strong, compact source of continuum radiation that ionizes the
material in the BLR. The nature of the continuum source and its
relation to the BLR and the jet remain unclear.  The continuum source
can be located in the accretion disk~\cite{lobanov:fr93} or in the hot
corona at 200-1000\,$R_\mathrm{g}$ above the accretion
disk~\cite{lobanov:fab04,lobanov:pon+04}.  A contribution from the jet
cannot be excluded~\cite{lobanov:fmr00}. The BLR is often assumed to
have an ellipsoidal shape, with at least a fraction of the BLR clouds
interacting with the jet plasma. There is also growing evidence for
the presence of a conical BLR component~\cite{lobanov:pop+01}
associated with a slower, sub-relativistic outflow originating in
outer regions of the accretion disk~\cite{lobanov:pop+01}.  The
sub-relativistic outflow is believed to be responsible also for the
broad absorption lines (BAL) observed in a number of
quasars~\cite{lobanov:elv00,lobanov:gal+04}. The disk, the BLR, and the
outflows must be closely connected, producing the bulk of the AGN
power. Reconstructing the physical mechanism behind this connection is
pivotal for understanding the AGN phenomenon in general.

\section{Zooming on the central engine}

The central engine, presumed to contain a SMBH, is an elusive
formation.  Direct detection and imaging of the event horizon of a
black hole cannot be done with present instruments.  The presence of
SMBH in galactic centres has been inferred so far only on the basis of
circumstantial evidence obtained from observations in the X-ray,
optical, near infrared, and radio domains.

\begin{figure}[t]
\includegraphics[width=0.95\textwidth,angle=0,bb=14 14 326 130,clip=true]{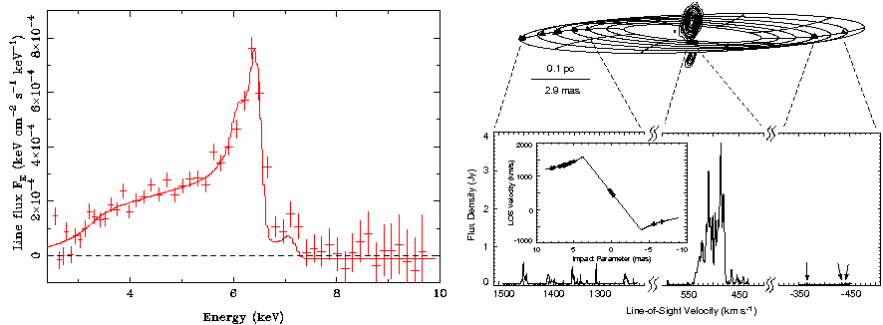}
\caption[]{{\bf Left:}~Relativistic Fe\,K$\alpha$ line profile in a
  Seyfert 1 galaxy MCG-6-30-15 obtained from a joint \emph{XMM-Newton}
  and \emph{Beppo-SAX} dataset~\cite{lobanov:fab+02}. The crosses mark the
  data points and the solid line marks the model in which the
  emission is generated in an accretion disk accretion and the inner
  radius of the emission is at $\approx 2\,R_\mathrm{g}$.  {\bf
    Right:}~H$_2$O masers in NGC\,4258. Top panel: actual maser
  positions (filled triangles and circles) superimposed on the
  radio-continuum emission (contours) and approximated by a model of
  warped disk. The filled square marks the best-fit location of the
  centre of the disk. Bottom panel: total spectrum of the maser
  emission and line of sigh velocities of individual spots fitted by a
  Keplerian disk model. The high-velocity masers trace a Keplerian
  curve to better than 1\%. Observations of H$_2$O masers in NGC\,4258
  have allowed to infer a geometric distance of $7.2\pm0.3$\,Mpc to
  the galaxy from the direct measurement of orbital motions in the
  maser spots. The motions imply a central object with a mass of
  $(3.9\pm0.1)\times 10^{7}\,\mathrm{M}_{\odot}$~\cite{lobanov:her+99}.}
\label{lobanov:fig4}
\end{figure}

\emph{X-ray observations} have revealed relativistically broadened line
profiles indicative of motions at speeds exceeding $10^5$\,km/s
(Fig.~\ref{lobanov:fig4}).  The X-ray emission generated above the
accretion disk interacts with the disk material and produces the iron
fluorescence
line~\cite{lobanov:fab+02,lobanov:fab+03,lobanov:min+03}. Observed line
profiles show relativistic speeds, Doppler shifts and
gravitational redshift.  Modelling of the line profiles constrains the
disk inclination and spin of the SMBH, which can be used to
distinguish between rotating and non-rotating black
holes~\cite{lobanov:zr04}.

\emph{Optical spectroscopy} has revealed the Doppler shift caused by
the fast Keplerian rotation of material in the accretion disk implying
masses of $1.5\times 10^9\,$M$_{\odot}$ in M\,84~\cite{lobanov:bow+98}
and $2.5\times 10^9\,$M$_{\odot}$ in M\,87~\cite{lobanov:mac+97}.

\emph{Near-infrared observations} of the nucleus of our own Galaxy
show the proper motion of stars
there~\cite{lobanov:eck+02,lobanov:ghe+00}. This motion implies a mass
of $\approx 3.7\times 10^6\,$M$_{\odot}$ enclosed within 45\,AU distance
from the Galactic Centre~\cite{lobanov:ghe+05}.

\emph{Radio observations} of maser lines in NGC\,4258
(Fig.~\ref{lobanov:fig4}) have yielded the most accurate measure of
distance and black hole mass in an external
galaxy~\cite{lobanov:her+99}.  High-resolution radio observations of
M\,87 have probed directly scales as small as
100\,$R_\mathrm{g}$~\cite{lobanov:jbl99,lobanov:kri+02}. The presence of
supermassive black holes in galactic centres is also implied
indirectly by the exceptional stability of jet
direction~\cite{lobanov:ohc89,lobanov:per+99} and apparent superluminal
motions that require highly relativistic
flows~\cite{lobanov:zen97,lobanov:zen+96}.

\subsection{Physics of relativistic outflows from SMBH}

\begin{figure}[t]
\begin{center}
\includegraphics[totalheight=0.95\textwidth,angle=-90,bb=14 14 181 326,clip=true]{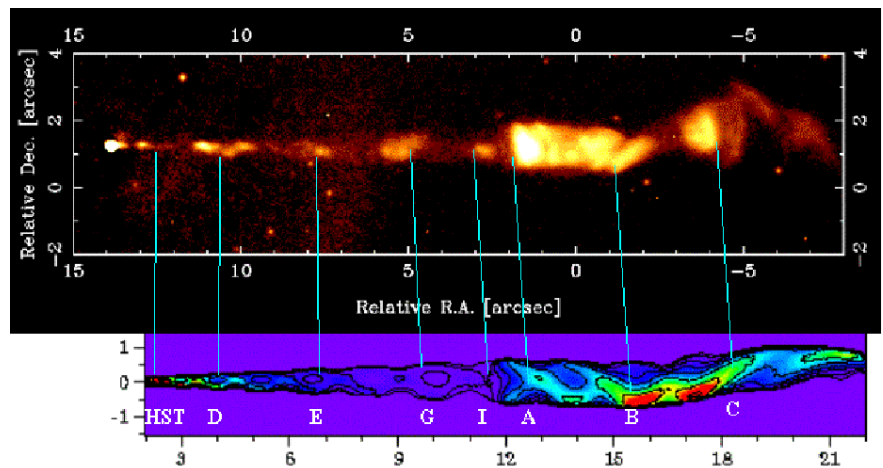}
\caption[]{Top: HST image of M\,87~\cite{lobanov:per+99}. Botom: Line-of-sight
synchrotron intensity image and contours at $\theta = 40^\circ$
calculated from an analytical model of the internal structure of the
jet using Kelvin-Helmholtz instability. Connecting lines
illustrate the correspondence between the two images. The model
represent well the brightness distribution in the jet on scales larger than 1\,kpc~\cite{lobanov:lhe03}. }
\label{lobanov:fig5}
\end{center}
\end{figure}

The activity of the central engine is accompanied by
highly-relativistic collimated outflows (jets) of plasma material
formed and accelerated in the vicinity of the black
hole~\cite{lobanov:fer98,lobanov:koi+02,lobanov:mku01}. Inhomogeneities
in the jet plasma appear as a series of compact radio knots (jet
components) observed on scales ranging from several light weeks to
about a kiloparsec~\cite{lobanov:zen97,lobanov:zen+96}. The kinematics
and spectral evolution of the knots are determined by relativistic
shocks~\cite{lobanov:lz99} and plasma
instabilities~\cite{lobanov:lob98b,lobanov:lz01,lobanov:lhe03}. Strong
relativistic shocks tend to dissipate on scales of $\sim
10$\,pc~\cite{lobanov:lz99} and internal structure of large-scale jets
is dominated by Kelvin-Helmholtz instability
(Fig.~\ref{lobanov:fig5}). Close to the central engine, the kinematic
and emission changes can reflect the dynamics in the central engine,
and this can be used effectively for estimating the properties of the
binary systems of SMBH~\cite{lobanov:ca04,lobanov:lr05}.

\section{SMBH in a larger context}

Supermassive black holes have now become the centrepiece of the modern
AGN paradigm. They are expected to form in the early Universe, in the
course of multiple mergers of dark matter halos (hierarchical
structure formation,~\cite{lobanov:kh00}).  SMBH are probably present
in all (or almost all) galaxies~\cite{lobanov:kr95,lobanov:mag+98} and
their masses are correlated with the velocity dispersion in galactic
bulges~\cite{lobanov:fm00,lobanov:geb+00,lobanov:mag+98,lobanov:tre+02}. SMBH
are closely related to just about every aspect of galactic activity,
and may be one of the most significant energy sources in the Universe.
Strength of the AGN activity depends critically on the dynamic
properties of gas and stars in the nuclear
region~\cite{lobanov:dok89,lobanov:dok91,lobanov:ips05}, and may be
connected to the presence of multiple SMBH in the host
galaxy~\cite{lobanov:lob05,lobanov:roo81,lobanov:roo85,lobanov:roo85b}.

Most of the energy in the vicinity of SMBH is created through
accretion, which ultimately releases the gravitational energy of
infalling matter and possibly mediates the process of releasing the
rotational energy of the SMBH itself~\cite{lobanov:ch00}. However, the
dominating mechanism and detailed physics of the energy release are
still not well understood --- it could be the ionizing continuum from the
accretion disk, the kinetic energy of the outflow~\cite{lobanov:bp82},
a magnetohydrodynamic mechanism~\cite{lobanov:mei01,lobanov:mku01} or
perhaps even a purely electromagnetic process~\cite{lobanov:bz76}.  A
detailed picture of physical conditions in the immediate vicinity of
SMBH is required for attempting to understand the mechanism for
production, conversion and transport of the gravitational energy of
the infalling material.

The most straightforward approach to studies of SMBH --- probing their
environment via direct imaging --- is still out of reach because of
extremely challenging requirements posed by this task on resolution
and dynamic range of observations~\cite{lobanov:bd05,lobanov:tak05}.
Direct detection and imaging of the event horizon of a black hole may
become possible with the future interferometric missions planned in
the X-ray~\cite{lobanov:cas05,lobanov:cas+00} and
radio~\cite{lobanov:zak+05} domains.

In this situation, physical properties of SMBH can be presently
constrained from detailed studies of their environment on scales
readily accessible for modern astrophysical instruments.  Most
important effects of SMBH on their environment are: 1)~strong
relativistic effects in the immediate vicinity of the event horizon,
providing characteristic distortion of background line and continuum
radiation~\cite{lobanov:bd05,lobanov:dov+04,lobanov:dkm04,lobanov:kvp92,lobanov:zr04,lobanov:tak05};
2)~processing of the galactic gas and shaping the mass distribution
and dynamics of gas and stars in the central regions of
galaxies~\cite{lobanov:ips05}; 3)~energy and matter release into the
ISM and IGM via outflows and emission (both substantially
non-isotropic) during galaxy formation and
evolution~\cite{lobanov:dsh05,lobanov:hs03}; 4)~stellar disruptions and
other recurrent events of energy release
(flares)~\cite{lobanov:hck04,lobanov:ncs04}.  Further detailed studies
of these effects will be instrumental for understanding the physics of
SMBH and their role in active galactic nuclei.

\section{Cosmological co-evolution of AGN and SMBH}

Recent years have seen an impressive progress in our understanding of
cosmological co-evolution of active galaxies and supermassive black
holes at their centres. The ``demographics'' of black holes can now be
studied in detail, using the celebrated
$M_\mathrm{bh}$--$\sigma_\mathrm{bulge}$
relation~\cite{lobanov:kin03,lobanov:tre+02} and its derivatives
connecting black hole masses in galaxies to observable properties of
their nuclear emission. Evidence accumulates for SMBH to be residing
practically in every galaxy with a nuclear
bulge~\cite{lobanov:hk00}. The black hole growth via accretion
correlates with the bulge growth via star
formation~\cite{lobanov:kh00}, which results in a tight correlation
between the masses of bulges and central black
holes~\cite{lobanov:hr04}.  The mass ratio
$M_\mathrm{bh}/M_\mathrm{bulge}$ is likely to be higher at $z=2$--3,
implying that black hole growth may induce the formation of galactic
bulges.  More massive galaxies and SMBH form at earlier cosmological
epochs~\cite{lobanov:has05}. Initial assembly of SMBH in most powerful
AGN begins at $z>10$~\cite{lobanov:bl03}, and primordial SMBH in AGN
are likely to contribute to reionization of the
Universe~\cite{lobanov:mad+04}. The cosmological growth of SMBH is
regulated by accretion, black hole mergers and stellar disruption
events. The evolutionary sequence of AGN can be roughly outlined
within four basic stages: manifestations of nuclear activity start
from extremely obscured objects with a prominent sub-millimetre
excess~\cite{lobanov:lut+00,lobanov:mb00}, then proceed to mainstream
QSO stage, followed by an extremely red object (ERO) stage, and
finally turning into ``normal'', ``inactive'' elliptical and spiral
galaxies~\cite{lobanov:dan+05}.

During the entire course of their evolution, AGN have a significant
impact on the interstellar and intergalactic
medium~\cite{lobanov:dsh05,lobanov:fab+03}. The radiative and kinetic
feedback from AGN deposits vast quantities of energy into the
environment on scales reaching several megaparsecs. The kinetic
feedback from jets and BAL outflows reaches energies of $\sim 0.01
\,M_\mathrm{bh}\, c^2$~\cite{lobanov:hei+05}. The radiative feedback
influences strongly the SMBH growth in
galaxies~\cite{lobanov:dsh05,lobanov:sdh05}.

Galactic mergers and binary black holes are probably pivotal for the
cosmological evolution of active galaxies.  Powerful AGN are most
likely produced in the course of galactic mergers~\cite{lobanov:hk00},
with every merger inducing two main episodes of accretion and star
formation: at the first contact and at the final coalescence of two
galaxies~\cite{lobanov:dsh05}. The nuclear activity is reduced when a
loss cone is formed and most of nuclear gas is accreted onto
SMBH~\cite{lobanov:dok91,lobanov:ips05}, but it can be maintained at a
relatively high level in the presence of a secondary black
hole~\cite{lobanov:dok89,lobanov:mer04}. It is possible that the nuclear
activity is closely related to dynamic evolution of binary SMBH in the
centres of active galaxies~\cite{lobanov:lob05}. Observational
investigations of binary black hole systems in galaxies will be
crucial for understanding the connection between SMBH and AGN.

\section{Fundamental questions}

Despite recent substantial advances in the field of AGN research,
nuclear activity in galaxies still holds a number of puzzling
questions that can only be answered with the next generation
instruments which would provide orders of magnitude increases in
sensitivity and resolution in just about every domain of astrophysical
observations.

Accretion of galactic gas, stars, and (possibly) dark matter on
supermassive black holes is now widely recognized as the most
plausible mechanism of maintaining the black hole growth and nuclear
activity in galaxies on cosmologically significant timescales. The
gravitational potential energy of the infalling material and the
rotational energy of the black hole are probably the primary sources
of the AGN power.  There is, however, a number of poorly understood
fundamental issues connecting the observed manifestations of nuclear
activity to the accretion process and physics of black holes. We still
do not know the exact mechanism by which the gravitational and
rotation energy is extracted and converted into emission and kinetic
energy of relativistic plasma ejected from the nucleus.  We need to
understand how the mechanism and efficiency of the energy release is
connected to the state of the SMBH (mass, spin, presence of a
secondary SMBH in the nucleus) and physical conditions in the nuclear
region. It is critical to understand whether and how these properties
differ in the objects with and without powerful outflows.  In a
broader scope, putting these properties in the cosmological context
would help attempting to reconstruct the pace of cosmological growth
of SMBH and the evolution of their environment.

Resolving these fundamental issues should provide a sufficient basis
for answering at least some of the ``eternal'' questions about
galactic activity. Is there a unified model for all of the different
types of AGN? Can we draw an evolutionary diagram for AGN?  What is
the relationship between the SMBH and the host galaxy? Do all galaxies
have SMBH? Were black hole seeds or by-products of galaxy formation?
How important is the feedback from supermassive black holes for the
structure formation in the Universe?  What is the role of accreting
black holes in the reionization of the Universe? What causes the
activity phase in a galaxy and sets its lifetime?  How are AGN jets
produced and collimated?  How do they interact with, and affect, the
host galaxy.  How is the SMBH activity connected to galaxy mergers and
central starbursts?  Are multiple SMBH common in AGN, and how do they
form and evolve? These are some of the problems that will hopefully be
understood and resolved in the next decades with the help of the new
generation of astrophysical facilities.

%

\end{document}